\documentclass[11pt,letterpaper]{article}

\usepackage[margin=1in]{geometry}
\usepackage[T1]{fontenc}
\usepackage{lmodern}
\usepackage{amsmath,amssymb}
\usepackage{graphicx}
\usepackage{tikz}
\usetikzlibrary{arrows.meta,backgrounds,fit,positioning}
\usepackage{longtable,booktabs,array,calc}
\usepackage{caption}
\usepackage{fancyvrb,upquote}
\usepackage{cite}
\usepackage{xurl}
\usepackage[hidelinks]{hyperref}

\newcommand{\topic}[1]{\subsubsection*{\normalsize #1}}

\title{GenAI-DBT: Evaluating a DBT-Informed Generative AI Platform for Mindfulness Skills Training: A Pilot Randomized Study}
\author{%
  Jingchen Ye\textsuperscript{\ensuremath{1,\dagger}}\qquad Yujia Chen\textsuperscript{\ensuremath{2,\dagger}}\qquad Leyang Han\textsuperscript{\ensuremath{1,\dagger}}\\[4pt]
  Siyuan Wang\textsuperscript{\ensuremath{3}}\qquad Mark Zachary Rosenthal\textsuperscript{\ensuremath{4,5}}\qquad Yucheng Jin\textsuperscript{\ensuremath{1,*}}\\[6pt]
  \parbox[t]{.97\textwidth}{\centering\footnotesize\textsuperscript{1}\,Division of Natural and Applied Sciences, Duke Kunshan University, \mbox{Kunshan, Jiangsu, China}}\\
  \parbox[t]{.97\textwidth}{\centering\footnotesize\textsuperscript{2}\,Department of Applied Psychology, School of Social and Behavioral Sciences, Nanjing University, \mbox{Nanjing, Jiangsu, China}}\\
  \parbox[t]{.97\textwidth}{\centering\footnotesize\textsuperscript{3}\,Department of Psychological Sciences, College of William and Mary, \mbox{Williamsburg, VA, United States}}\\
  \parbox[t]{.97\textwidth}{\centering\footnotesize\textsuperscript{4}\,Department of Psychiatry and Behavioral Sciences, Duke University Medical Center, \mbox{Durham, NC, United States}}\\
  \parbox[t]{.97\textwidth}{\centering\footnotesize\textsuperscript{5}\,Department of Psychology and Neuroscience, Duke University, \mbox{Durham, NC, United States}}\\[5pt]
  \small \textsuperscript{$\dagger$}\,These authors contributed equally.\\[2pt]
  \small \textsuperscript{*}\,Corresponding author: Yucheng Jin (\href{mailto:yucheng.jin@dukekunshan.edu.cn}{\nolinkurl{yucheng.jin@dukekunshan.edu.cn}})%
}
\date{}
\hypersetup{pdftitle={GenAI-DBT: Evaluating a DBT-Informed Generative AI Platform for Mindfulness Skills Training: A Pilot Randomized Study},
  pdfauthor={Jingchen Ye, Yujia Chen, Leyang Han, Siyuan Wang, Mark Zachary Rosenthal, Yucheng Jin}}

\begin{document}
\maketitle
\vspace{-1.8em}

\begin{abstract}
\textbf{Background:} Dialectical behavior therapy (DBT) is an evidence-based approach for developing emotion regulation skills, but applying these skills in everyday situations can be difficult without structured guidance. Generative artificial intelligence (AI) may support DBT skills practice through personalized and context-sensitive guidance. However, what it adds beyond static web resources or general-purpose large language model (LLM) chatbots remains unclear.

\textbf{Objective:} This study evaluated the feasibility, acceptability, and engagement of three digital formats for DBT skills training. We also explored preliminary emotion-related outcomes.

\textbf{Methods:} In this three-arm randomized controlled pilot study, 45 university students were assigned to GenAI-DBT, a general-purpose LLM chatbot (Chatbot), or a static web-based resource (Web). Forty-two participants completed the 30-day study. Outcomes included the Mindful Attention Awareness Scale (MAAS), Difficulties in Emotion Regulation Scale (DERS), System Usability Scale (SUS), objective engagement metrics, session-level self-reported mood, and textual sentiment derived from user-authored messages. Nine participants also completed semi-structured interviews.

\textbf{Results:} Forty-two of 45 participants completed the study. SUS scores were within acceptable ranges across all three conditions, with no significant between-group differences. GenAI-DBT had the highest observed participant-week adherence rate (71.4\%), compared with Web (53.3\%) and Chatbot (42.3\%); these engagement differences were descriptive. MAAS and DERS showed no significant time-by-group interactions, and self-reported mood changes did not differ significantly across conditions. Among the two conversational conditions, GenAI-DBT showed a larger positive shift in textual sentiment than Chatbot (mean change 4.28 vs 1.59; $P=.02$; Cohen $d=0.99$). This measure reflects linguistic valence rather than emotional state. Interview participants valued contextualized and scenario-based skills practice, while also identifying response latency, repetition, and limited personalization as areas for improvement.

\textbf{Conclusions:} GenAI-DBT was feasible and acceptable for 30-day DBT mindfulness skills training. Differences across formats were more apparent in engagement and interaction-level responses than in broader self-reported outcomes. These findings do not establish clinical efficacy, but they support further evaluation of DBT-informed conversational AI in larger and more clinically relevant samples.
\end{abstract}

\section{Introduction}\label{introduction}

Emotion dysregulation is a transdiagnostic feature of many mental health conditions and is associated with increased psychological distress, impaired functioning, and elevated risk of self-harm and suicidality \cite{ref-1}. Dialectical behavior therapy (DBT) is one of the most widely used evidence-based treatments for emotion dysregulation and was originally developed for individuals with Borderline Personality Disorder (BPD) and chronic suicidal behaviors \cite{ref-2,ref-3}.

DBT comprises four core skills modules: mindfulness, distress tolerance, emotion regulation, and interpersonal effectiveness. Among these, mindfulness skills help individuals observe, describe, and engage with present-moment experiences in a nonjudgmental manner, thereby increasing awareness of emotional and cognitive processes while reducing automatic reactivity \cite{ref-6,ref-7,ref-8}. Existing studies suggest that DBT mindfulness skills can improve attentional awareness, reduce anxiety, and alleviate symptoms of emotional distress, depression, and hopelessness \cite{ref-9,ref-10,ref-11}. 

Regular skills practice is an important component of DBT, as using DBT skills in daily life is central to transferring what is learned in training to emotionally challenging situations \cite{ref-12,ref-13,ref-14}. However, opportunities for continued skills training and support are often limited outside structured therapy settings, and individuals may struggle to generalize and apply learned skills independently when facing emotionally challenging situations \cite{ref-15}. To address these barriers, digital mental health interventions have increasingly been used to extend psychoeducation and skills training beyond face-to-face settings. These approaches can improve accessibility and scalability, but often provide limited personalization or interactive guidance \cite{ref-16,ref-17,ref-18}.

Generative artificial intelligence (AI), particularly large language models (LLMs), introduces a more interactive approach to digital skills training. Compared with static digital interventions, LLM-based systems can engage users in natural conversations, adapt responses to individual circumstances, and provide context-sensitive guidance during skills practice \cite{ref-19,ref-21}. However, general-purpose LLM chatbots are not specifically designed around evidence-based DBT curricula and may lack structured skills training and guidance aligned with DBT principles \cite{ref-22}.

To address this gap, we developed GenAI-DBT and compared three digital delivery formats: a static web-based platform (Web), a general-purpose LLM chatbot (Chatbot), and GenAI-DBT, which combines conversational interaction with structured DBT guidance, personalized feedback, and scenario-based practice. In this design, Web provides standardized content, Chatbot adds conversational interaction, and GenAI-DBT further incorporates DBT-informed structure and contextualized practice.

This pilot study examined the feasibility and acceptability of three digital formats for DBT skills training, with particular attention to user engagement. We also explored preliminary emotion-related outcomes. University students were randomized to Web, Chatbot, or GenAI-DBT. We assessed pre- and post-intervention outcomes, platform engagement, session-level self-reported mood and textual sentiment, and post-intervention user feedback. The three-arm design allowed us to examine what structured and personalized DBT guidance may add beyond conversational AI alone.

Beyond evaluating feasibility and acceptability, this study introduces GenAI-DBT, a DBT-informed conversational platform that combines personalized, context-aware skills coaching with scenario-based practice. Comparing GenAI-DBT with both a static Web platform and a general-purpose chatbot also helps distinguish what conversational interaction and DBT-specific structure may contribute to digital skills training.

\subsection{Related Work}\label{related-work}

\topic{Digital Interventions for DBT Skills Training}
\label{digital-interventions-for-dbt-skills-training}

Digital technologies have long been used to expand access to evidence-based mental health interventions. Early approaches relied largely on web-based psychoeducation and internet-delivered cognitive behavioral therapy (iCBT), demonstrating that psychological interventions could be delivered at scale outside traditional face-to-face settings \cite{ref-23,ref-24}. Access, however, is only part of the problem. Sustaining engagement remains difficult, particularly when support is largely self-guided and cannot readily adapt to users' changing needs \cite{ref-29,ref-30}.

Digital DBT interventions have followed a similar trajectory. Early applications such as DBT Coach demonstrated the feasibility of providing in-the-moment coaching for specific skills, including Opposite Action \cite{ref-26}. More recent mobile applications have expanded this approach through features such as diary cards, broader skills content, and support alongside comprehensive DBT programs \cite{ref-27}. Yet moving DBT skills onto a digital platform does not by itself solve the challenge of applying them in emotionally difficult situations. Usability, sustained engagement, and real-world skill use remain important concerns across digital mental health interventions \cite{ref-28,ref-29,ref-30}.

More interactive approaches have therefore begun to incorporate conversational interfaces, immediate feedback, and scenario-based practice \cite{ref-21,ref-31,ref-32,ref-38}. These systems move beyond information delivery by allowing users to respond, practice, and receive guidance during an interaction. However, many remain general-purpose or offer limited adaptation to users' emotional contexts and learning progress. This leaves room for systems that combine the accessibility of digital delivery with more personalized, structured, and context-aware DBT skills practice.

\topic{Conversational AI for Mental Health and DBT}
\label{conversational-ai-for-mental-health-and-dbt}

Recent research has increasingly explored conversational agents for mental health support, ranging from rule-based chatbots to large language model (LLM)-based systems \cite{ref-19,ref-31}. Compared with static web-based platforms, conversational systems allow users to interact in natural language and receive immediate, responsive feedback. Studies of mental health chatbots have reported promising findings for feasibility, acceptability, engagement, and symptom-related outcomes, although the evidence base remains heterogeneous \cite{ref-17,ref-36,ref-37,ref-38}.

General-purpose LLM chatbots offer greater conversational flexibility, but they are not specifically designed around evidence-based therapeutic curricula. Their responses may therefore vary in quality and may not provide the structure needed for systematic skills acquisition and practice \cite{ref-19,ref-22}. This distinction is particularly relevant to DBT. Skills are not learned through information alone; they require repeated practice and application in relevant situations \cite{ref-2,ref-6}.

For DBT mindfulness training, conversational AI creates an opportunity to move from explaining a skill to practicing it in context. LLMs can support context-aware dialogue, personalized feedback, and interactive rehearsal \cite{ref-19,ref-20,ref-21}. Yet most existing work has focused on general mental health support or broader therapeutic conversations. Empirical work on generative AI specifically designed for structured DBT mindfulness skills training remains limited, and direct comparisons with existing digital delivery formats are particularly scarce \cite{ref-19,ref-36,ref-37,ref-38}. This gap motivated the development and evaluation of GenAI-DBT.

\topic{Supporting Evidence-Based Mental Health Interventions with LLMs}
\label{supporting-evidence-based-mental-health-interventions-with-llms}

Recent studies have explored the use of large language models (LLMs) for psychoeducation, emotional support, symptom monitoring, and therapeutic conversations. As these systems move into mental health settings, an important challenge is keeping their responses reliable and aligned with the intended intervention. Approaches such as retrieval-augmented generation (RAG), structured prompting, and hybrid architectures have therefore been used to ground model responses in external knowledge and place clearer constraints on generation \cite{ref-36,ref-40}. Much of this work has focused on response quality, reliability, and safety. Less attention has been paid to a different question: how these systems can support the repeated learning and practice of therapeutic skills over time.

\section{Methods}\label{methods}

\subsection{Materials}\label{materials}

\topic{Participant Data}\label{participant-data}

Anonymized participant-level and session-level data were collected from 45 university students who participated in one of three DBT-based digital interventions: Web, Chatbot, and GenAI-DBT. The dataset included pre- and post-intervention questionnaire responses assessing mindfulness (Mindful Attention Awareness Scale, MAAS) and emotion regulation difficulties (Difficulties in Emotion Regulation Scale, DERS), as well as post-intervention ratings of system usability using the System Usability Scale (SUS), together with platform-generated usage logs. These logs contained timestamped mood-tracker entries for all three conditions and user-authored text records for the interactive conditions (Chatbot and GenAI-DBT). For session-level analyses, each daily usage episode was indexed by a unique session-group identifier (\texttt{group\_id}), allowing mood entries, dialogue turns, and calendar date to be linked within the same analytic unit.

\topic{Sentiment Analysis Resource}\label{sentiment-analysis-resource-1}

Textual records were analyzed using a transformer-based semantic sentiment pipeline informed by prior psychotherapy research on automated sentiment measurement \cite{ref-35}. We used the multilingual XLM-R architecture \cite{ref-41} with the publicly available terrencewee12/xlm-roberta-base-sentiment-multilingual-finetuned-v2 checkpoint \cite{ref-42}. For each user-authored message, the model estimated negative, neutral, and positive class probabilities. These probabilities were mapped to a 0-10 expected-value score as 0 x P(negative) + 5 x P(neutral) + 10 x P(positive). The resulting score represents linguistic valence in the submitted text and was not treated as a direct measure of the user\textquotesingle s underlying emotional or clinical state.

\subsection{Intervention Platform}\label{intervention-platform}

\topic{Overview}\label{overview}

\begin{figure}[!htbp]
\centering
\resizebox{\linewidth}{!}{%
\begin{tikzpicture}[
  font=\small,
  >={Stealth[length=1.8mm]},
  node distance=6mm and 7mm,
  box/.style={draw, rounded corners=2pt, align=center, inner sep=3pt,
              minimum height=9mm, text width=31mm, fill=white},
  core/.style={draw, rounded corners=2pt, align=center, inner sep=3pt,
               minimum height=13mm, text width=43mm, fill=black!10, very thick},
  support/.style={draw, rounded corners=2pt, align=center, inner sep=3pt,
                  minimum height=8mm, text width=39mm, fill=white},
  io/.style={align=center, font=\footnotesize\itshape},
  arr/.style={-{Stealth[length=1.8mm]}, semithick},
  darr/.style={-{Stealth[length=1.8mm]}, semithick, dashed}
]

\node[io] (input) {User\\Input};
\node[box, right=of input] (llm) {\textbf{Shared LLM Layer}\\
  \footnotesize Natural-language conversation};
\node[core, right=of llm] (dbt) {\textbf{DBT-Informed Design}\\
  \footnotesize Structured prompting and curated DBT knowledge\\[1pt]
  \scriptsize DBT Coach Test embedded for\\scenario-based skills assessment};
\node[box, right=of dbt] (output) {\textbf{Personalized Output}\\
  \footnotesize Skills guidance and feedback};

\draw[arr] (input) -- (llm);
\draw[arr] (llm) -- (dbt);
\draw[arr] (dbt) -- (output);

\node[support, below=17mm of dbt, xshift=-48mm] (resources)
  {\textbf{Online Resources}\\\footnotesize Text and video instruction};
\node[support, right=of resources] (mood)
  {\textbf{Mood Tracker}\\\footnotesize Explicit self-report logging};
\node[support, right=of mood] (engagement)
  {\textbf{Secondary Engagement}\\\textbf{Cues}\\
   \footnotesize Progress and achievement};

\begin{scope}[on background layer]
  \node[fit=(input)(llm)(dbt)(output),
        draw=black!45, dashed, rounded corners, fill=black!4, inner sep=3.5mm,
        label={[font=\footnotesize\bfseries]above:
          Conversational DBT Skills Workflow}] {};
  \node[fit=(resources)(mood)(engagement),
        draw=black!45, dashed, rounded corners, fill=black!4, inner sep=3mm,
        label={[font=\footnotesize\bfseries]below:
          Additional Platform Components}] {};
\end{scope}

\end{tikzpicture}%
}
\caption{Structure of the GenAI-DBT platform. The DBT Coach Test is embedded within DBT Informed Design for LLM, while achievement cues are shown as secondary engagement support.}
\label{fig:genai-dbt-architecture}
\end{figure}
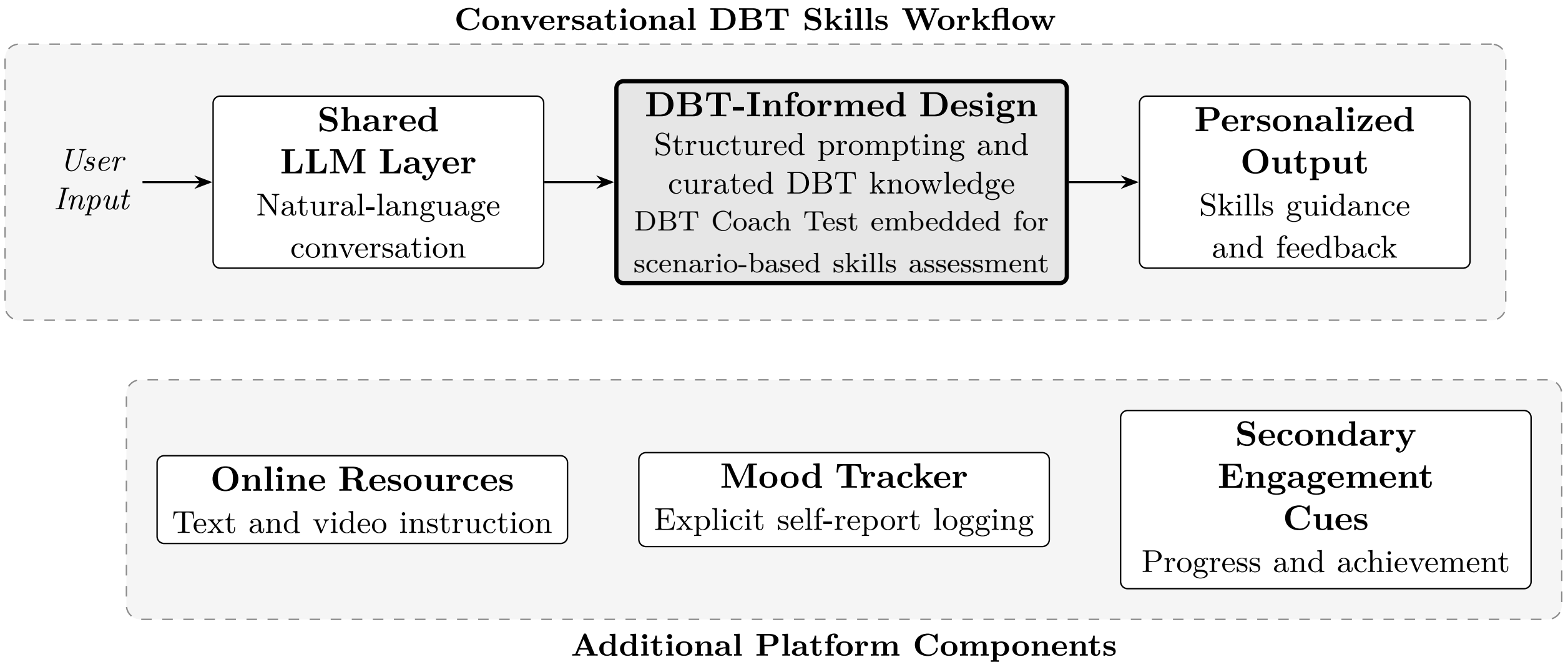

GenAI-DBT was implemented as a web-based platform for DBT mindfulness skills training. Its four primary components were online resources, a mood tracker, a shared LLM-Based Chatbot, and DBT Informed Design for LLM. Within the DBT-informed component, a scenario-based DBT Coach Test supported skills application and structured feedback. Progress and achievement cues served as secondary engagement supports rather than a standalone primary component. User progress, mood logs, and interaction records were stored under anonymized identifiers on access-controlled infrastructure. Figure~\ref{fig:genai-dbt-architecture} shows the resulting architecture.

\topic{Online Resources}\label{online-resources}

The online resources feature provided DBT skills teaching videos that could be viewed directly on the platform. Participants could learn the target skills through video-based instruction without additional interaction, representing a conventional digital format for skills delivery. This feature served as the platform's lowest-interactivity mode of learning and allowed participants to review core DBT content at their own pace.

\topic{Mood Tracker}\label{mood-tracker}

\begin{figure}[!htbp]
\centering
\includegraphics[width=.62\linewidth]{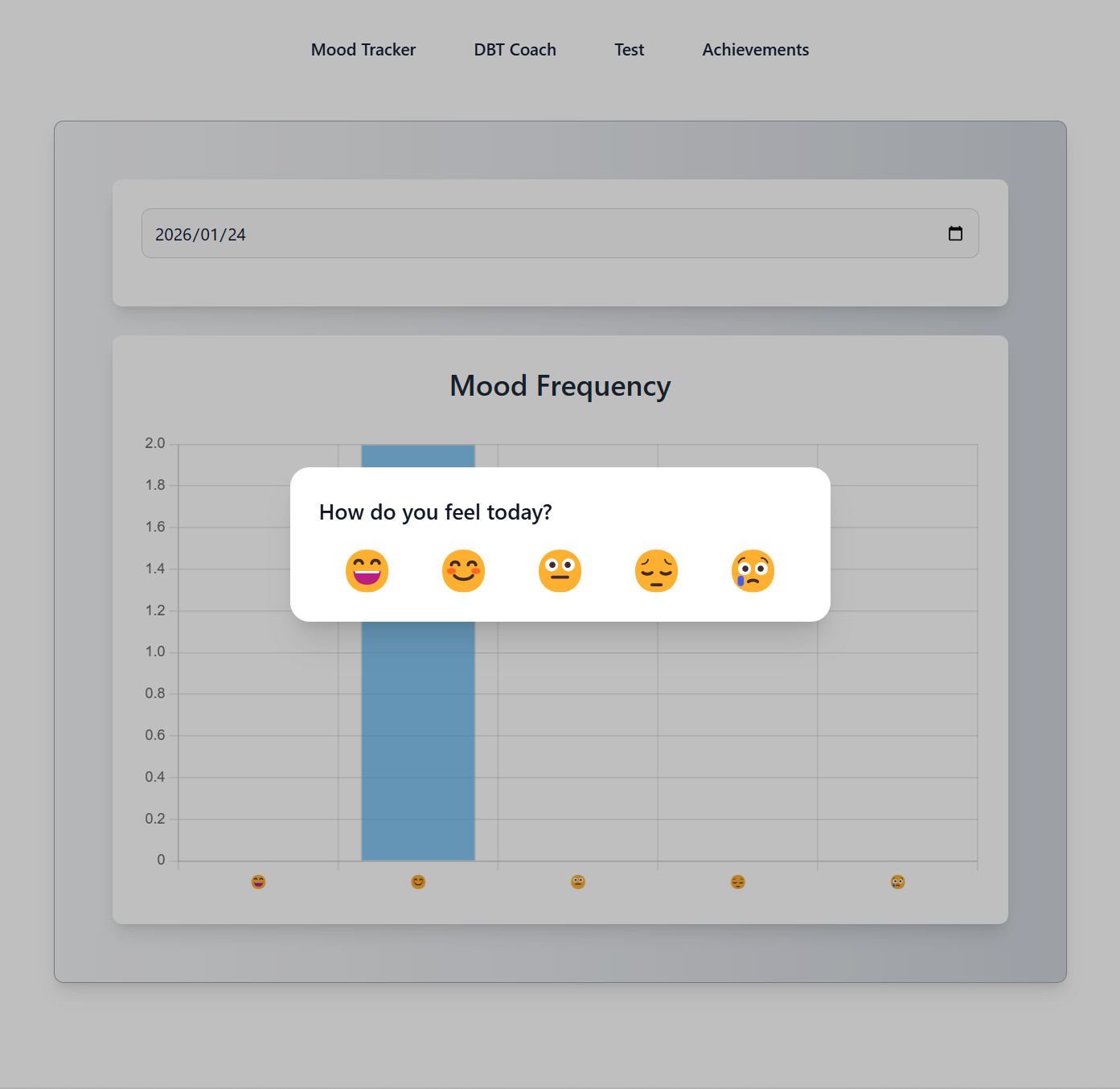}
\caption{Emoji-based mood tracker interface used for session-level mood assessment.}
\label{fig:mood-tracker}
\end{figure}

The mood tracker presented five emojis representing affective states ranging from sadness to joy. Each selection was automatically logged, and the user\textquotesingle s daily mood pattern was displayed as a bar chart. Participants could access this feature at any time. The system also prompted users before and after skills practice to record their mood, allowing calculation of within-session change in explicit self-reported affect (Figure~\ref{fig:mood-tracker}).

\topic{LLM-Based Chatbot}\label{llm-based-chatbot-1}

\begin{figure}[!htbp]
\centering
\includegraphics[width=.73\linewidth]{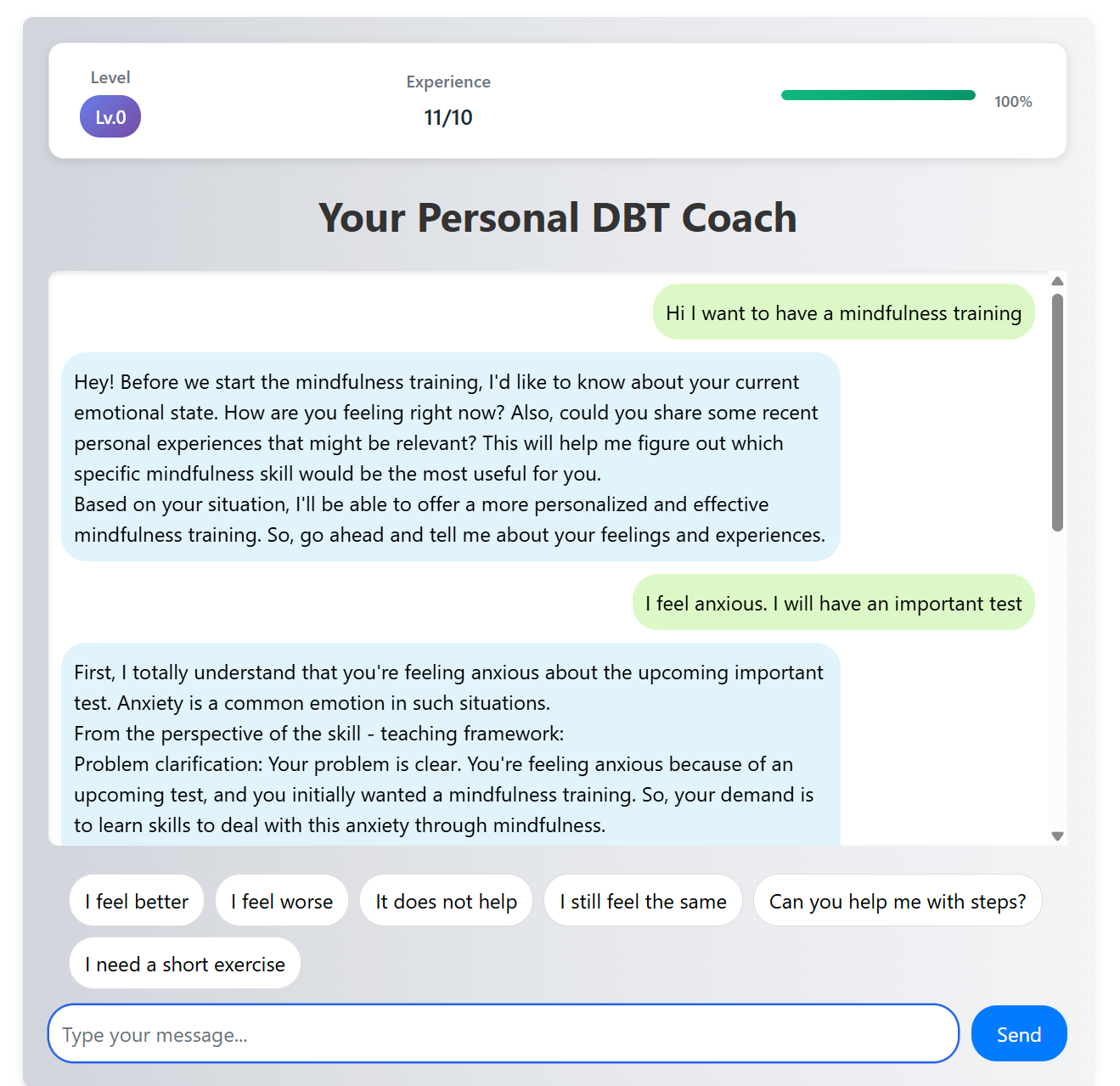}
\caption{LLM-based chatbot training interface example.}
\label{fig:chatbot-interface}
\end{figure}

The LLM-Based Chatbot provided the platform's shared conversational layer for the two LLM-enabled conditions. At this level, the system's function was to support basic natural-language interaction, allowing users to describe their current situation, ask questions, and receive immediate conversational responses in a neutral, non-judgmental tone. This layer supplied the underlying dialogue capability shared by the Chatbot and GenAI-DBT conditions, but it did not by itself define how responses were structured according to DBT principles. An example of the LLM-based chatbot training interface is shown in Figure~\ref{fig:chatbot-interface}.

\topic{DBT Informed Design for LLM}\label{dbt-informed-design-for-llm-1}

DBT Informed Design for LLM was the structured instructional layer available only in the GenAI-DBT condition. It combined DBT-specific prompt engineering with retrieval from a curated DBT knowledge base to organize personalized skills practice. The English DBT coaching prompt is provided in Appendix~\ref{app:dbt-prompt}. The workflow identified a relevant DBT skills domain, collected brief contextual information, generated a practice context, guided application of the selected skill, and provided structured feedback. A scenario-based DBT Coach Test was embedded within this workflow to assess skills application using personalized situations and corrective feedback. Progress and achievement cues were used only as secondary supports for continued practice.

Together, the four primary components and their embedded supports constituted the full GenAI-DBT platform. The experimental conditions differed in which components were available, allowing comparison of static resources, general conversational interaction, and the added DBT-informed instructional layer.

\subsection{Data Collection and Procedure}\label{data-collection-and-procedure}

The study was conducted over a 30-day period using a remote, asynchronous data collection method.

\begin{itemize}
\item
  Participant onboarding: upon recruitment, participants provided informed consent and completed baseline demographic questionnaires. They were then introduced to the GenAI-DBT interface, where the system collected initial personalization data to calibrate future interactions.
\item
  Interaction data: the system logged user-agent interactions, including daily assessment scores, engagement metrics (frequency and completion rates), and anonymized dialogue logs for qualitative analysis of adherence to DBT principles and user responses.
\item
  Privacy and ethics: interaction logs were de-identified, and the system included a safety guardrail that prohibited storage or generation of sensitive personal data outside the scope of the immediate scenario-based interaction.
\end{itemize}

\subsection{Study Design}\label{study-design}

\topic{Participants}\label{participants}

Participants were undergraduate university students recruited through online advertisements distributed via social media platforms (e.g., WeChat Moments and Xiaohongshu). Inclusion criteria were: (a) aged between 18 and 22 years, (b) currently enrolled as a full-time university student, (c) ownership of a smartphone or computer with stable internet access, and (d) willingness to participate in a 30-day intervention and complete baseline and post-intervention assessments.

Exclusion criteria included elevated risk for depression or suicidality, assessed during prescreening using the Patient Health Questionnaire-9 (PHQ-9). Participants scoring above the predefined cutoff (PHQ-9 \(>\) 14, indicating moderately severe depression) or endorsing the suicidality item (item 9 \(\geq\) 1) were excluded and provided with appropriate mental health resources \cite{ref-43}. A total of 45 participants met inclusion criteria and completed baseline assessments. Participants were randomly assigned to one of three experimental conditions (Web, Chatbot, or GenAI-DBT). During the intervention period, three participants discontinued participation for personal or scheduling reasons, resulting in a final per-protocol sample of 42 participants.

\topic{Intervention Design}\label{intervention-design}

This study employed a three-arm randomized controlled pilot design to compare modes of DBT mindfulness skills delivery. Across all three conditions, participants received identical core content covering Wise Mind; the what skills (Observe, Describe, Participate); and the how skills (Nonjudgmentally, One-mindfully, Effectively). Materials included text explanations, brief videos, guided exercises, and structured reflection prompts. Practice recommendations were consistent across conditions (at least four sessions per week, approximately 10 minutes per session). Conditions differed in delivery mode and interactivity rather than instructional content.

\topic{Experimental Conditions}\label{experimental-conditions}

\begin{itemize}
\item
  Web (baseline): static, web-based DBT mindfulness materials (written text and prerecorded video content) with basic mood tracking and reflection functions.
\item
  Chatbot: the same core materials as Web-DBT with an LLM-based chatbot allowing participants to ask questions and receive conversational responses.
\item
  GenAI-DBT: the shared chatbot was supplemented with DBT-specific prompt engineering, retrieval-augmented generation, personalized scenario practice, an embedded application assessment, structured feedback, and secondary progress cues.
\end{itemize}

The comparison conditions represented two common modes of digital support. Web delivered structured DBT mindfulness content through static pages and prerecorded media \cite{ref-18,ref-25}. Chatbot added a general-purpose LLM conversational layer, reflecting the growing use of conversational agents for mental health information and support \cite{ref-22}. GenAI-DBT combined this conversational capability with the DBT-informed instructional workflow described above. These contrasts therefore separated static content delivery, general conversational interaction, and the added DBT-informed design. Table~\ref{tab:features} summarizes the available components.

{\small
\setlength{\LTpre}{.65\baselineskip}
\begin{longtable}{@{}
  >{\raggedright\arraybackslash}p{(\linewidth - 6\tabcolsep) * \real{.52}}
  >{\centering\arraybackslash}p{(\linewidth - 6\tabcolsep) * \real{.13}}
  >{\centering\arraybackslash}p{(\linewidth - 6\tabcolsep) * \real{.15}}
  >{\centering\arraybackslash}p{(\linewidth - 6\tabcolsep) * \real{.20}}@{}}
\caption{Platform-supported features by intervention group.}\label{tab:features}\\
\toprule
Feature & Web & Chatbot & GenAI-DBT \\
\midrule
\endfirsthead
\toprule
Feature & Web & Chatbot & GenAI-DBT \\
\midrule
\endhead
\midrule
\endfoot
\bottomrule
\endlastfoot
Online resources & {\checkmark} & {\checkmark} & {\checkmark} \\
Mood tracker & {\checkmark} & {\checkmark} & {\checkmark} \\
LLM-Based Chatbot & & {\checkmark} & {\checkmark} \\
DBT Informed Design for LLM (personalized skills guidance, scenario-based application assessment, and structured feedback) & & & {\checkmark} \\
\end{longtable}
}

\topic{Procedure}\label{procedure-1}

After completing online screening and providing informed consent, eligible participants completed baseline questionnaires assessing mindfulness (MAAS) and emotion regulation (DERS). Participants were then randomly assigned to one of the three experimental conditions. An online onboarding session was conducted on Day 0, during which participants received instructions on how to use their assigned system and completed an initial guided mindfulness practice. The intervention period lasted 30 consecutive days (Day 1--30). Participants were instructed to engage in mindfulness practice at least four times per week, with each session lasting approximately 10 minutes. Brief reflection entries were submitted following each practice session.

Post-intervention assessments were conducted between Day 31 and Day 35 and included the MAAS, DERS, SUS, and two questions about DBT skills. A convenience subsample of nine participants (three from each condition) subsequently took part in individual semi-structured interviews. Participation was voluntary, and the first three volunteers from each condition were enrolled. Interviews were conducted in Chinese via VooV, lasted approximately 15--20 minutes, and were audio-recorded and transcribed for analysis.

\topic{Measures}\label{measures}

\textbf{Mindfulness:} Mindfulness was assessed using the Mindful Attention Awareness Scale (MAAS), a widely used self-report measure of dispositional mindfulness \cite{ref-44}. The MAAS consists of 15 items rated on a 6-point Likert-type scale, with higher total scores indicating greater mindful attention and awareness. The MAAS was administered at baseline and post-intervention.

\textbf{Emotion Regulation:} Emotion regulation difficulties were measured using the Difficulties in Emotion Regulation Scale (DERS) \cite{ref-45}. The Chinese version used in this study consisted of 32 items rated on a 5-point Likert scale, with higher scores reflecting greater difficulties in emotion regulation \cite{ref-46}. The DERS was administered at baseline and post-intervention.

System Usability: Perceived usability was evaluated after the intervention using the 10-item System Usability Scale (SUS) \cite{ref-47}. Items were rated on a 5-point Likert scale and transformed to a total score from 0 to 100, with higher scores indicating more favorable perceived usability. Empirical SUS benchmarking was used to interpret the observed score ranges \cite{ref-48}.

Engagement: Objective engagement was derived from the existing session-group identifiers in the platform logs for the per-protocol sample (Web n=15, Chatbot n=13, GenAI-DBT n=14). Metrics included active days, total practice episodes, mean practice episodes per complete week, weekly target attainment, and sustained activity across the 30-day observation period. The target was at least four session groups in each complete 7-day block (Days 1-7, 8-14, 15-21, and 22-28); Days 29-30 were described separately and excluded from weekly adherence rates. For Chatbot and GenAI-DBT, message-level logs also provided user turns, total user messages, user-authored text volume, and active conversation duration.

Proximal Self-Reported Mood and Textual Sentiment: Participants recorded momentary affect before and after practice using an emoji-based mood tracker mapped to a 0-10 scale (sad/crying=0, slightly sad=2.5, neutral=5.0, slightly positive/happy=7.5, very happy=10). Within each session group, change in self-reported mood was the final minus initial mood rating. For Chatbot and GenAI-DBT, change in textual sentiment was the final minus initial XLM-R score across user-authored messages in the same group. Mood represents explicit self-reported affect; textual sentiment represents linguistic valence in submitted text. These constructs were analyzed separately. Web did not generate comparable dialogue and therefore contributed only to mood analyses.

\subsection{Data Analysis}\label{data-analysis}

\topic{Quantitative Data}\label{quantitative-data}

Changes in mindfulness (MAAS) and emotion regulation difficulties (DERS) were examined using mixed-design analyses of variance (ANOVA), with time (pre vs post) as a within-subject factor and group (Web, Chatbot, GenAI-DBT) as a between-subject factor. Where appropriate, analyses of covariance (ANCOVA) were conducted to control for baseline differences. Within-group changes were further examined using paired-samples \(t\) tests. System usability (SUS) scores were summarized descriptively and compared across groups using one-way ANOVA.

Planned contrasts compared (1) Chatbot with Web to characterize the addition of a shared conversational layer and (2) GenAI-DBT with Chatbot to characterize the additional DBT-informed design, including DBT-specific prompting and retrieval-augmented generation.

Session-level platform data were linked using the existing group\_id, defined as a platform-recorded daily usage episode. No alternative session regrouping was introduced for the engagement analysis.

For self-reported mood, the first and last emoji records within a session group were used to calculate delta mood. For textual sentiment, the first and last user-authored messages were scored on the 0-10 XLM-R expected-value scale and used to calculate delta sentiment.

Because mood and textual sentiment capture different constructs, their estimates were reported independently and were not combined into a single summary indicator.

Engagement metrics were computed at the participant, participant-week, participant-day, and session levels as appropriate. Active days were unique calendar dates with at least one session group. Total practice episodes were counts of existing group\_id values. Participant-week adherence was the proportion of four complete weeks with at least four recorded session groups.

For conversational sessions, active duration was the sum of time intervals between adjacent user messages when the interval was 120 minutes or less. Longer intervals were treated as pauses and excluded; single-user-turn conversations contributed 0 minutes. Conversation metrics were not applicable to Web.

To address repeated sessions within participants, session-group mood and textual-sentiment changes were averaged within participant before between-condition testing. Two-sided permutation tests with 5,000 resamples were used for participant-level contrasts, and Cohen\textquotesingle s d was reported for continuous outcomes. Engagement outcomes were summarized descriptively using means and SDs together with medians and IQRs.

All proximal response findings were interpreted according to their measurement construct and the pilot, per-protocol design; they were not treated as evidence of clinical efficacy.

\topic{Human Validation of Transformer-Based Textual Sentiment Scoring}\label{human-validation-of-transformer-based-textual-sentiment-scoring}

We selected 34 of 340 valid textual-sentiment session groups (10.0\%) for independent human validation. Sampling was proportional by condition and intervention phase (Days 1-10, 11-20, and 21-30), yielding 14 Chatbot and 20 GenAI-DBT session pairs. The first and last user-authored message from each selected group were exported, for 68 messages in total.

Two independent raters who could evaluate both Chinese and English rated linguistic sentiment on the same 0-10 scale. Raters were blind to condition, study day, original first/last position, XLM-R score, and each other\textquotesingle s ratings; within-pair message order was randomized. Human-human reliability was assessed using a two-way random-effects, absolute-agreement, single-measure ICC (ICC(2,1)) with a 95\% bootstrap CI \cite{ref-49}. XLM-R agreement with the mean human score was summarized by absolute-agreement ICC, Spearman correlation, and mean absolute error (MAE). Agreement for first-to-last delta sentiment was evaluated using the same statistics and directional agreement.

\topic{Qualitative Data}\label{qualitative-data}

Semi-structured interview transcripts were analyzed in NVivo using a hybrid deductive--inductive thematic analysis approach \cite{ref-50,ref-51}. The analysis focused on participants' experiences of using their assigned platform, including engagement, perceived usefulness, usability, application of DBT skills in daily life, perceived limitations, and suggestions for improvement. The Enlight framework proposed by Baumel et al. \cite{ref-52} informed the initial areas of attention, particularly engagement, usability, and intervention quality, but was used as a sensitizing framework rather than a fixed coding scheme. This allowed themes specific to AI-supported DBT skills training to also emerge from the interview data.

Two psychology researchers independently read and coded the transcripts. They subsequently compared their coding and discussed differences until consensus was reached. Codes were revised and grouped into broader themes through repeated reference to the original transcripts. Both favorable and critical experiences were retained in the analysis so that the themes reflected not only perceived benefits but also barriers to use and areas requiring further platform refinement. Once the thematic structure was established, coded material was compared across the Web, Chatbot, and GenAI-DBT conditions to identify shared and condition-specific patterns.

Representative quotations were selected to illustrate the final themes. Quotations were translated from Chinese into English by a researcher. Generative AI was used to assist with language checking of the translated quotations, after which the researcher reviewed the English wording against the original Chinese transcripts to preserve the intended meaning.

\subsection{Ethical considerations}\label{ethical-considerations}

Ethical approval was obtained from the IRBs of Duke Kunshan University (2025LH098) and Nanjing University (NJUPSY202506015). All participants provided informed consent. Data were anonymized and securely stored. Participants received RMB 100 for study completion, and interview participants received a Starbucks gift card of equivalent value.

\section{Results}\label{results}

\subsection{Quantitative Findings}\label{quantitative-findings}

\topic{Participant Flow and Sample Characteristic}\label{participant-flow-and-sample-characteristic}

A total of \(N = 45\) university students met inclusion criteria and completed baseline assessments. Participants were randomly assigned to Web (\(n = 15\)), Chatbot (\(n = 15\)), or GenAI-DBT (\(n = 15\)). During the 30-day intervention, \(n = 3\) participants discontinued due to personal or scheduling reasons. The final per-protocol sample included \(N = 42\) participants who completed both pre- and post-intervention assessments and were included in the primary quantitative analyses (Web: \(n = 15\), Chatbot: \(n = 13\), GenAI-DBT: \(n = 14\)).

Baseline demographic characteristics and questionnaire scores are presented in Table~\ref{tab:baseline}. The mean age of participants was 20.60 years (SD = 1.75), and 73.8\% identified as female. No substantial baseline differences were observed across groups.

{\small
\setlength{\LTpre}{.65\baselineskip}
\begin{longtable}{@{}
  >{\raggedright\arraybackslash}p{(\linewidth - 6\tabcolsep) * \real{.31}}
  >{\centering\arraybackslash}p{(\linewidth - 6\tabcolsep) * \real{.22}}
  >{\centering\arraybackslash}p{(\linewidth - 6\tabcolsep) * \real{.23}}
  >{\centering\arraybackslash}p{(\linewidth - 6\tabcolsep) * \real{.24}}@{}}
\caption{Participant characteristics and baseline measures by intervention group.}\label{tab:baseline}\\
\toprule
Variable & Web-DBT (\(n = 15\)) & Chat-DBT (\(n = 13\)) & GenAI-DBT (\(n = 14\)) \\
\midrule
\endfirsthead
\toprule
Variable & Web-DBT (\(n = 15\)) & Chat-DBT (\(n = 13\)) & GenAI-DBT (\(n = 14\)) \\
\midrule
\endhead
\midrule
\endfoot
\bottomrule
\endlastfoot
Age, mean (SD) & 20.93 (2.28) & 20.54 (1.45) & 20.29 (1.38) \\
Gender, \(n\) (\%) & & & \\
Female & 10 (66.7\%) & 10 (76.9\%) & 11 (78.6\%) \\
Male & 5 (33.3\%) & 3 (23.1\%) & 3 (21.4\%) \\
PHQ-9, mean (SD) & 11.93 (1.44) & 11.31 (1.60) & 10.86 (1.86) \\
\end{longtable}
}

\topic{Mindfulness and Emotion Regulation Self-Report Measures (MAAS and DERS)}\label{mindfulness-and-emotion-regulation-self-report-measures-maas-and-ders}

Changes in mindfulness (MAAS) and emotion regulation difficulties (DERS) were examined using mixed-design ANOVA with time (pre vs. post) as the within-subjects factor and group (Web, Chatbot, GenAI-DBT) as the between-subjects factor.

For MAAS, there was no significant main effect of time, \(F(1,39) = 1.71\), \(p = .20\), no main effect of group, and no time \(\times\) group interaction, \(F(2,39) = 0.62\), \(p = .54\). An ANCOVA controlling for baseline MAAS scores similarly indicated no significant group differences at post-intervention, \(F(2,38) = 0.21\), \(p = .81\), partial \(\eta^{2} = .011\).

For DERS, a significant overall effect of time was observed, F(1,39)=6.74, p=.01, indicating lower reported emotion-regulation difficulty at post-intervention across the pooled sample. Neither the group effect nor the time-by-group interaction was significant, F(2,39)=0.05, p=.95. ANCOVA controlling for baseline DERS likewise showed no post-intervention group difference, F(2,36)=0.06, p=.95, partial eta squared=.015. Thus, the time effect does not identify a condition-specific advantage.

Exploratory within-group paired-sample tests did not identify statistically significant pre-post changes for MAAS or DERS within any individual condition. Descriptively, mean MAAS increased in Chatbot and GenAI-DBT and mean DERS decreased in all conditions, but the absence of group interactions limits condition-specific interpretation. Table~\ref{tab:maas-ders} reports the observed means and SDs.

{\small
\setlength{\LTpre}{.65\baselineskip}
\begin{longtable}{@{}
  >{\raggedright\arraybackslash}p{(\linewidth - 8\tabcolsep) * \real{.13}}
  >{\raggedright\arraybackslash}p{(\linewidth - 8\tabcolsep) * \real{.10}}
  >{\centering\arraybackslash}p{(\linewidth - 8\tabcolsep) * \real{.25}}
  >{\centering\arraybackslash}p{(\linewidth - 8\tabcolsep) * \real{.25}}
  >{\centering\arraybackslash}p{(\linewidth - 8\tabcolsep) * \real{.27}}@{}}
\caption{Pre- and post-intervention MAAS and DERS scores by group.}\label{tab:maas-ders}\\
\toprule
Measure & Time & Web-DBT mean (SD) & Chat-DBT mean (SD) & GenAI-DBT mean (SD) \\
\midrule
\endfirsthead
\toprule
Measure & Time & Web-DBT mean (SD) & Chat-DBT mean (SD) & GenAI-DBT mean (SD) \\
\midrule
\endhead
\midrule
\endfoot
\bottomrule
\endlastfoot
MAAS & Pre & 62.00 (13.26) & 55.92 (17.31) & 60.00 (11.52) \\
MAAS & Post & 61.67 (10.89) & 61.46 (13.82) & 63.36 (13.83) \\
MAAS & \(\Delta\) & -0.33 & 5.54 & 3.36 \\
DERS & Pre & 86.27 (20.05) & 87.31 (18.71) & 82.21 (18.96) \\
DERS & Post & 80.00 (20.28) & 78.85 (17.05) & 75.07 (16.69) \\
DERS & \(\Delta\) & -6.27 & -8.46 & -7.14 \\
\end{longtable}
}

\topic{Usability and Preliminary Acceptability (SUS)}\label{usability-and-preliminary-acceptability-sus}

Perceived usability was assessed after the intervention using the SUS. Mean scores were 69.50 (SD 14.12) for Web, 76.73 (SD 13.56) for Chatbot, and 73.39 (SD 12.50) for GenAI-DBT. All three scores fell within ranges generally characterized as acceptable in empirical SUS benchmarking \cite{ref-48}. Chatbot had the highest observed mean, followed by GenAI-DBT and Web, but the group difference was not statistically significant and was not interpreted as equivalence. Considered with 42 of 45 participants completing the study and the objective engagement patterns reported below, the results provide preliminary support for usability and acceptability rather than clinical efficacy. Item-level and total scores are shown in Table~\ref{tab:sus}.

{\small
\setlength{\LTpre}{.65\baselineskip}
\begin{longtable}{@{}
  >{\raggedright\arraybackslash}p{(\linewidth - 6\tabcolsep) * \real{.47}}
  >{\centering\arraybackslash}p{(\linewidth - 6\tabcolsep) * \real{.17}}
  >{\centering\arraybackslash}p{(\linewidth - 6\tabcolsep) * \real{.18}}
  >{\centering\arraybackslash}p{(\linewidth - 6\tabcolsep) * \real{.18}}@{}}
\caption{System Usability Scale (SUS) scores by intervention group.}\label{tab:sus}\\
\toprule
Question & Web-DBT mean (SD) & Chat-DBT mean (SD) & GenAI-DBT mean (SD) \\
\midrule
\endfirsthead
\toprule
Question & Web-DBT mean (SD) & Chat-DBT mean (SD) & GenAI-DBT mean (SD) \\
\midrule
\endhead
\midrule
\endfoot
\bottomrule
\endlastfoot
I think I would like to use the system frequently. & 2.73 (0.88) & 3.00 (0.71) & 2.64 (0.84) \\
I found the system to be unnecessarily complex. & 1.80 (1.27) & 1.69 (1.60) & 1.57 (1.22) \\
I thought the system was easy to use. & 3.20 (0.94) & 3.31 (0.75) & 3.14 (0.86) \\
I think that I would need support of a technical person to be able to use the system. & 1.93 (1.16) & 0.77 (0.73) & 1.14 (1.10) \\
I found the various functions in the system were well integrated. & 2.40 (0.91) & 2.85 (0.90) & 2.86 (0.77) \\
I thought there was too much inconsistency in the system. & 1.07 (0.88) & 0.92 (0.86) & 1.14 (0.86) \\
I would imagine that most people would learn to use the system very quickly. & 3.53 (0.64) & 3.46 (0.52) & 3.50 (0.52) \\
I found the system very cumbersome to use. & 1.07 (1.28) & 1.08 (0.95) & 1.07 (0.73) \\
I felt very confident using the system. & 3.00 (0.93) & 3.23 (0.83) & 3.00 (0.88) \\
I needed to learn a lot of things before I could get going with the System. & 1.20 (1.27) & 0.69 (0.95) & 0.86 (0.95) \\
System Usability Scale total score & 69.50 (14.12) & 76.73 (13.56) & 73.39 (12.5) \\
\end{longtable}
}

\topic{30-Day Engagement Outcomes}\label{day-engagement-outcomes}

Across the per-protocol sample, the platform logs contained 547 practice episodes: 189 for Web, 152 for Chatbot, and 206 for GenAI-DBT. Mean active days were 12.60 (SD 4.73), 11.69 (SD 6.38), and 14.36 (SD 3.95), respectively. Mean total practice episodes followed the same ordering because Web and Chatbot contributed at most one session group per active day, whereas a small number of GenAI-DBT days contained more than one episode.

Across the four complete 7-day weeks, the participant-week adherence rate for the target of at least four practice episodes was 53.3\% for Web, 42.3\% for Chatbot, and 71.4\% for GenAI-DBT. The proportion of participants meeting the target in all four complete weeks was 40.0\% (6/15), 23.1\% (3/13), and 50.0\% (7/14), respectively. Days 29-30 were excluded from these adherence rates; only one additional GenAI-DBT episode occurred during those two days. These descriptive patterns indicate sustained use by a subset of each condition, with the highest observed weekly adherence in GenAI-DBT.

Conversation-level engagement was available for 144 Chatbot and 203 GenAI-DBT sessions. Median user turns per conversation were 3 (IQR 2-6) for Chatbot and 7 (IQR 5-10) for GenAI-DBT. After excluding pauses longer than 120 minutes, median active conversation duration was 3.84 minutes (IQR 0.64-9.16) and 5.13 minutes (IQR 2.48-10.33), respectively. These interaction metrics characterize how the two conversational formats were used and were not interpreted as measures of clinical response.

{\small
\setlength{\LTpre}{.65\baselineskip}
\begin{longtable}{@{}
  >{\raggedright\arraybackslash}p{(\linewidth - 6\tabcolsep) * \real{.42}}
  >{\centering\arraybackslash}p{(\linewidth - 6\tabcolsep) * \real{.18}}
  >{\centering\arraybackslash}p{(\linewidth - 6\tabcolsep) * \real{.19}}
  >{\centering\arraybackslash}p{(\linewidth - 6\tabcolsep) * \real{.21}}@{}}
\caption{Thirty-day objective engagement metrics by intervention group.}\label{tab:engagement}\\
\toprule
\textbf{Metric} & \textbf{Web} & \textbf{Chatbot} & \textbf{GenAI-DBT} \\
\midrule
\endfirsthead
\toprule
\textbf{Metric} & \textbf{Web} & \textbf{Chatbot} & \textbf{GenAI-DBT} \\
\midrule
\endhead
\midrule
\endfoot
\bottomrule
\endlastfoot
Active days, mean (SD) & 12.60 (4.73) & 11.69 (6.38) & 14.36 (3.95) \\
Total practice episodes, mean (SD) & 12.60 (4.73) & 11.69 (6.38) & 14.71 (4.03) \\
Practice episodes per complete week, mean (SD) & 3.15 (1.18) & 2.92 (1.60) & 3.66 (1.00) \\
Participant-week adherence, \% & 53.3 & 42.3 & 71.4 \\
Met target in all four complete weeks, \% & 40.0 & 23.1 & 50.0 \\
Conversation sessions, n & N/A & 144 & 203 \\
User turns per conversation, median (IQR) & N/A & 3 (2-6) & 7 (5-10) \\
User messages per participant, mean (SD) & N/A & 46.3 (40.2) & 117.9 (61.2) \\
User-authored characters per participant, mean (SD) & N/A & 1608.0 (1373.9) & 1928.2 (1354.2) \\
Active conversation duration, median minutes (IQR) & N/A & 3.84 (0.64-9.16) & 5.13 (2.48-10.33) \\
\end{longtable}
}

\topic{Immediate Self-Reported Mood}\label{immediate-self-reported-mood}

Explicit within-session change in self-reported mood was assessed using the emoji scale. Participant-level mean delta mood was 0.611 for GenAI-DBT, 0.307 for Web, and 0.144 for Chatbot. This numerical ordering describes a proximal self-report pattern; the between-group differences were not statistically significant.

\begin{figure}[!htbp]
\centering
\includegraphics[width=.92\linewidth]{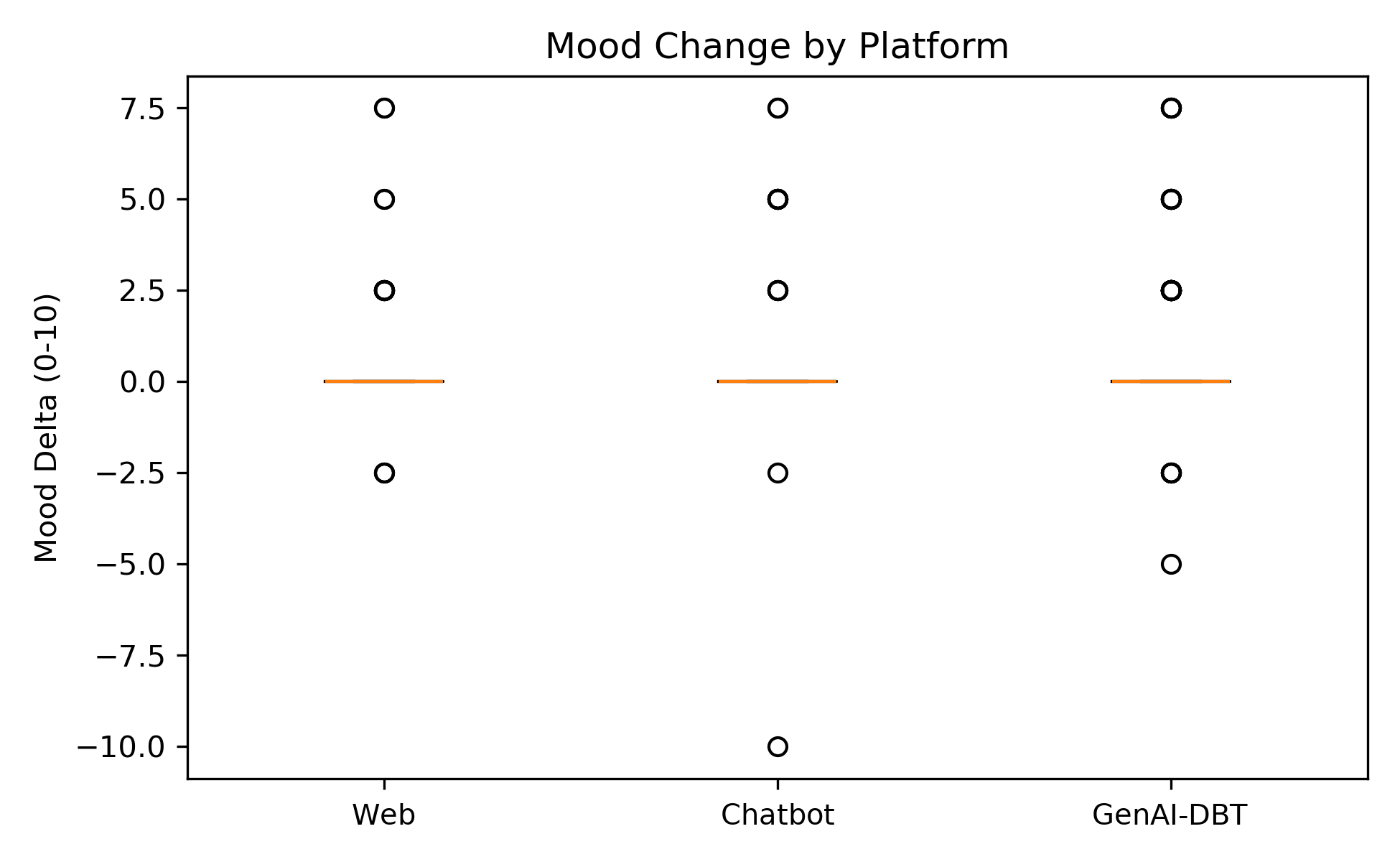}
\caption{Participant-level change in explicit self-reported mood by platform. Horizontal lines show medians (all 0); open circles show non-zero participant-level values.}
\label{fig:mood-change}
\end{figure}

Two-sided permutation tests with 5,000 resamples did not identify a significant GenAI-DBT difference relative to Chatbot (p=.23, d=0.49) or Web (p=.21, d=0.48). The observed direction is therefore reported as exploratory rather than evidence of superior mood response (Figure~\ref{fig:mood-change}).

\topic{Human Validation of Textual Sentiment Scoring}\label{human-validation-of-textual-sentiment-scoring}

Across the 68 independently rated messages, human-human absolute agreement was ICC(2,1) = 0.800 (95\% bootstrap CI 0.709-0.870). Agreement between XLM-R scores and the mean human ratings was ICC=0.820 (95\% bootstrap CI 0.645-0.952), Spearman rho=0.796, and MAE=0.974 points.

For first-to-last delta sentiment across the 34 session pairs, model-human agreement was ICC=0.872, Spearman rho=0.842, MAE=1.251, and directional agreement=82.4\%. These validation statistics were used to calibrate interpretation of the automated textual-sentiment findings.

\topic{Immediate Textual Sentiment}\label{immediate-textual-sentiment}

\begin{figure}[!htbp]
\centering
\includegraphics[width=.92\linewidth]{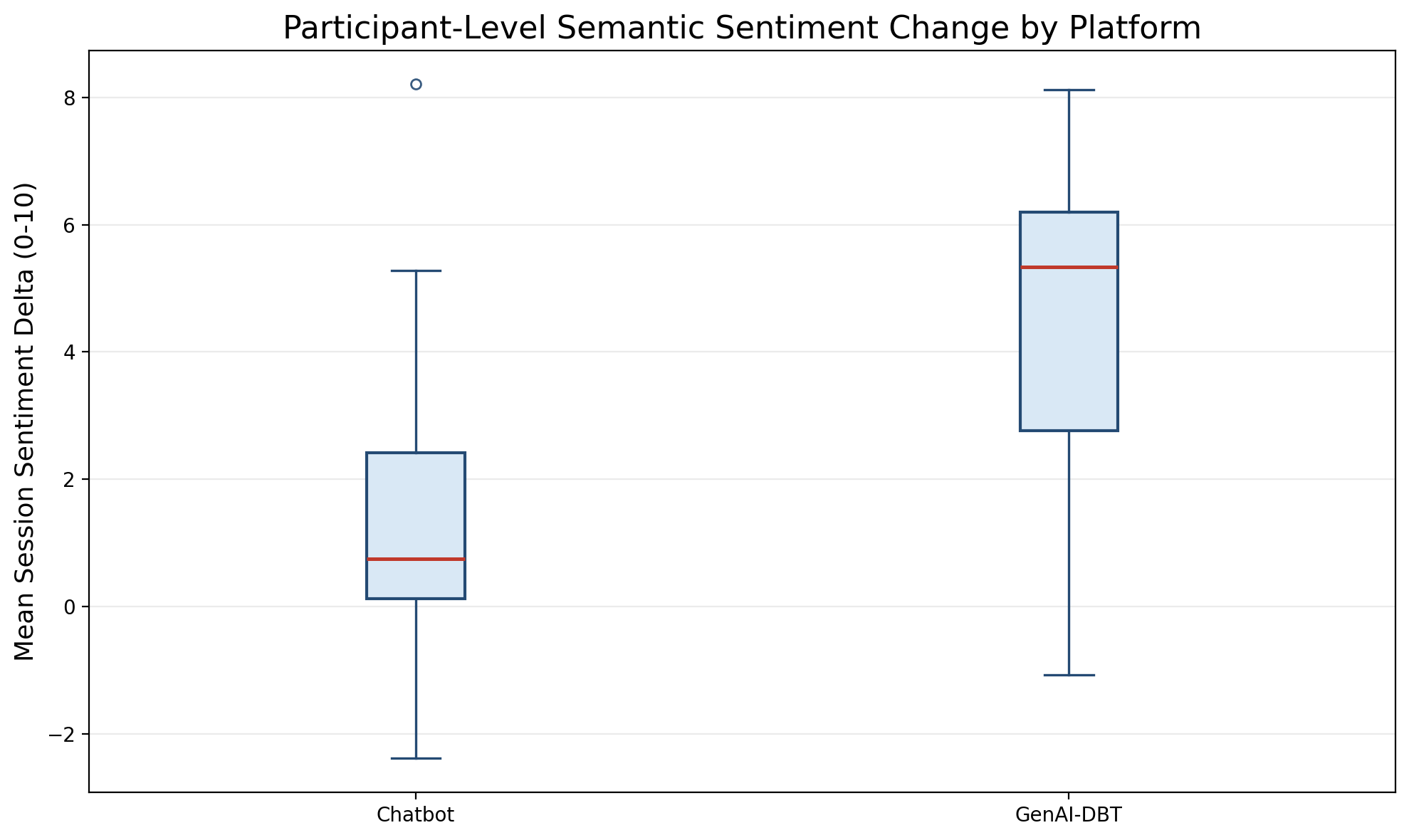}
\caption{Participant-level change in XLM-R-derived textual sentiment for the two conversational conditions.}
\label{fig:sentiment-change}
\end{figure}

Textual sentiment was evaluated for Chatbot and GenAI-DBT because Web did not generate comparable dialogue. Mean participant-level delta sentiment was 4.282 for GenAI-DBT and 1.588 for Chatbot. The difference was statistically significant in the two-sided permutation test (p=.02, d=0.99). This finding indicates a larger positive shift in the linguistic valence of user-authored text in GenAI-DBT and is consistent with a contribution of the DBT-informed design beyond the shared chatbot layer. It does not establish a corresponding change in underlying emotional state or clinical outcome (Figure~\ref{fig:sentiment-change}).

\topic{30-Day Engagement and Proximal Outcome Trajectories}\label{day-engagement-and-proximal-outcome-trajectories}

\begin{figure}[!htbp]
\centering
\includegraphics[width=.98\linewidth]{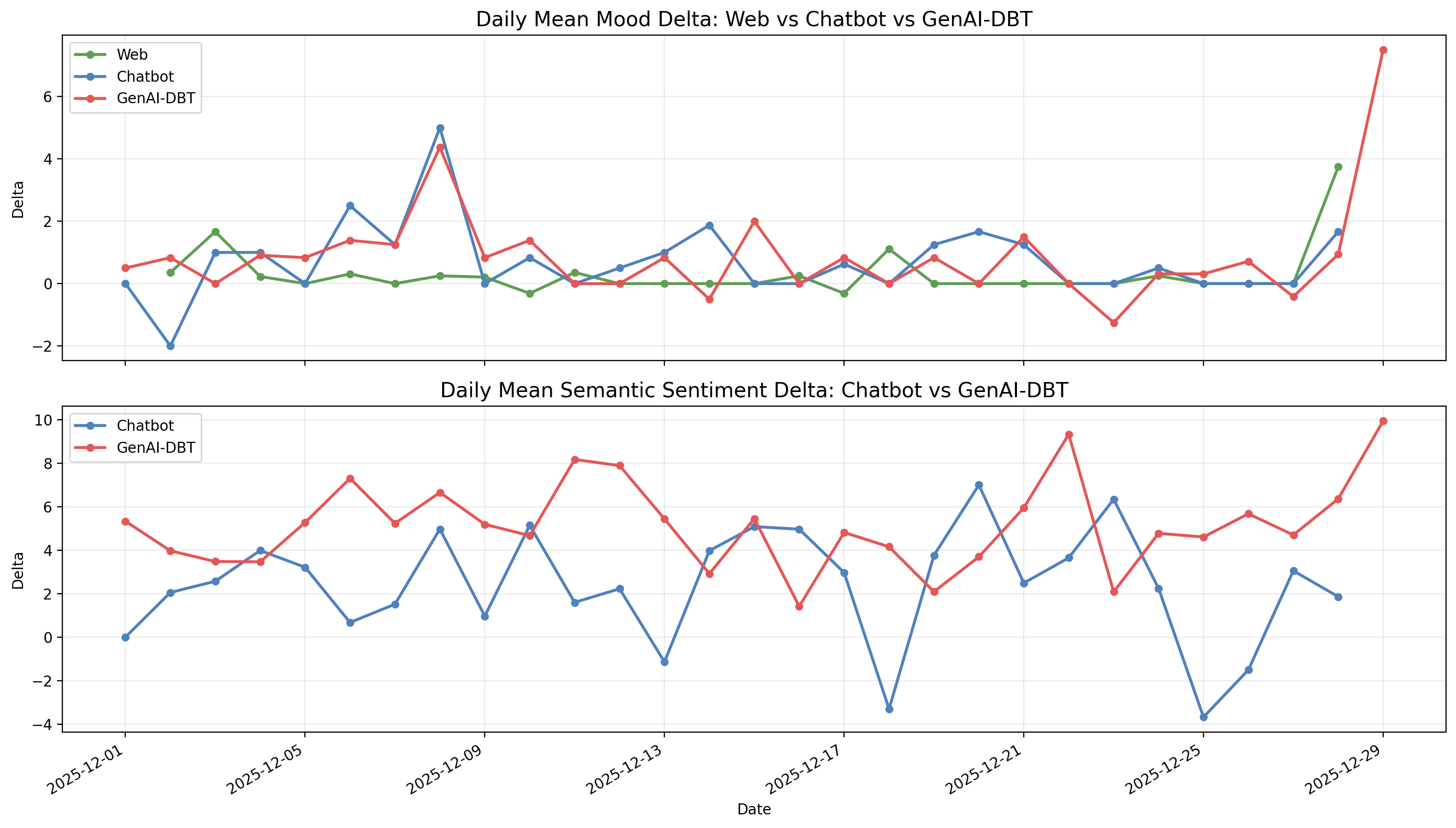}
\caption{Daily proximal response trajectories. The top panel shows day-level mean change in explicit self-reported mood; the bottom panel shows day-level mean change in XLM-R-derived textual sentiment.}
\label{fig:proximal-trajectories}
\end{figure}

Among session groups with valid mood data, the descriptive trajectory analysis included 189 Web, 87 Chatbot, and 178 GenAI-DBT groups. Mean daily delta mood was 0.301 across 27 days with valid Web mood data, 0.711 across 28 Chatbot days, and 0.894 across 29 GenAI-DBT days; positive daily means occurred on 40.7\%, 53.6\%, and 65.5\% of those days, respectively. Valid XLM-R textual-sentiment trajectories included 144 Chatbot and 196 GenAI-DBT groups. Mean daily delta sentiment was 2.379 for Chatbot and 5.171 for GenAI-DBT, with positive daily means on 82.1\% and 100.0\% of observed days. These day-level values describe patterns within the 30-day observation window and do not establish persistence beyond that period (Figure~\ref{fig:proximal-trajectories}).

\begin{figure}[!htbp]
\centering
\includegraphics[width=.82\linewidth]{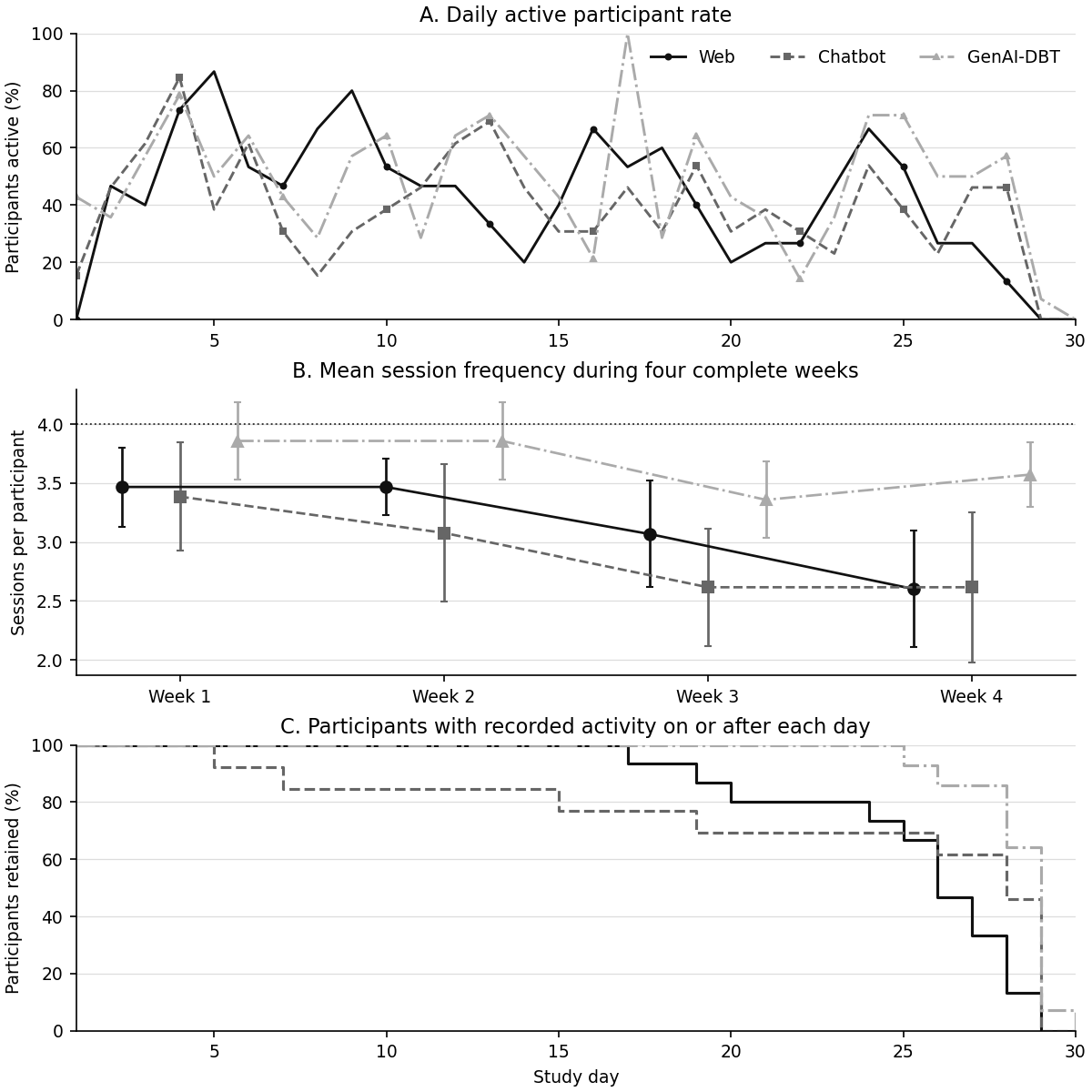}
\caption{Thirty-day objective engagement trajectories. Panel A shows the daily active-participant rate. Panel B shows mean practice episodes per participant during the four complete 7-day weeks; error bars are standard errors and the dotted line marks the four-session target. Panel C shows the proportion of participants with recorded activity on or after each study day.}
\label{fig:engagement-trajectories}
\end{figure}

The objective engagement trajectories complement the proximal response trajectories by showing when participants returned to the systems and how practice frequency changed across the four complete weeks. Figure~\ref{fig:engagement-trajectories} presents daily active-participant rates, weekly practice frequency, and the proportion of participants with recorded activity on or after each study day. The curves are descriptive and do not imply treatment persistence after the 30-day observation period.

\subsection{Qualitative Findings}\label{qualitative-findings}

Analysis of the nine semi-structured interviews yielded five themes concerning engagement, perceived usefulness, application of DBT skills, boundaries of AI-supported emotional support, and usability. The themes reflected both common experiences across the three conditions and features that participants associated with particular modes of delivery.

\topic{Theme 1: Engagement and integration into daily use}\label{theme-1-engagement-and-integration-into-daily-use}

Participants described different ways of incorporating the three platforms into daily practice. In the Web condition, engagement depended largely on self-initiation. Some participants found it difficult to think of using static materials when they were already experiencing strong emotions. As one participant explained, \emph{``When it was mostly text, I wouldn't really think of using it when I felt emotional.''} [Participant A1]. Another noted that regular use would have been difficult without external prompting: \emph{``If I had to rely on myself to remember to use it, I probably wouldn't\ldots{} unless there were reminders.''} [Participant A3].

Participants in the Chatbot condition described a more active form of engagement. Some adapted the way they communicated with the system in order to obtain responses that felt more relevant. One participant reflected, \emph{``I realized that how I ask questions really matters\ldots{} when I described my thoughts and feelings in more detail, the responses became much more helpful\ldots{} so I started to actively adjust the way I interacted with it.''} [Participant B2]. Another described the chatbot as something they might turn to when feeling stuck, while distinguishing this from dependence: \emph{``When I feel confused or unable to solve problems, I would think of using it first\ldots{} but I wouldn't say I depend on it.''} [Participant B3].

In the GenAI-DBT condition, participants also mentioned structured features that encouraged continued practice. In particular, achievement and experience systems gave some participants a clearer sense of progress. One participant stated, \emph{``The achievement and experience system makes me more motivated to complete the tasks.''} [Participant C1], while another said, \emph{``The achievement system makes me feel like I've reached a goal\ldots{} it motivates me to keep using it.''} [Participant C2].

These accounts suggest that the way participants engaged with the intervention depended not only on the therapeutic content, but also on how easily the platform could be incorporated into everyday use.

\topic{Theme 2: Perceived usefulness and personalization}\label{theme-2-perceived-usefulness-and-personalization}

Perceived usefulness was closely related to whether the content felt understandable, responsive, and relevant to the participant's situation. In the Web condition, some participants felt that static educational materials did not provide enough explanation for users who were unfamiliar with DBT. One participant commented, \emph{``For someone who has never been exposed to DBT\ldots{} the explanations and teaching could be more detailed\ldots.''} [Participant A1].

Participants in the Chatbot condition often emphasized the combination of emotional acknowledgment and concrete suggestions. One participant described the interaction as less ``machine-like'' than expected because the chatbot both acknowledged emotions and offered practical strategies: \emph{``It doesn't feel like other AI\ldots{} it doesn't have that strong `machine-like' feeling\ldots{} it approaches my problems from two sides---first by acknowledging my emotions, and second by giving me concrete methods\ldots{} like breathing or calming down.''} [Participant B1]. Another similarly noted, \emph{``It felt emotionally supportive, and it actually gave me concrete methods based on my state like breathing or grounding rather than just abstract suggestions.''} [Participant B2].

At the same time, greater structure did not always translate into greater personalization. A participant in the GenAI-DBT condition reported that responses could become repetitive across different situations: \emph{``Sometimes I ask different questions, but the solutions it gives are quite similar\ldots{} I wish it could be more specific.''} [Participant C3]. This was an important limitation because participants appeared to value guidance most when it felt closely connected to the situation they had described.

\topic{Theme 3: Applying DBT skills in everyday situations}\label{theme-3-applying-dbt-skills-in-everyday-situations}

Participants also differed in how they described using DBT skills outside the platform. In the Web condition, some participants found that accessing and working through static materials required more effort than they were willing to invest during emotionally intense moments. One participant explained, \emph{``In daily life situations\ldots{} it didn't really help me apply the skills. When I experienced emotions, I didn't have the patience to go through the process of clicking and recording\ldots{} it felt inconvenient in those moments.''} [Participant A1].

Participants in the Chatbot condition described using suggested strategies more directly in daily situations. One participant reported following suggestions such as breathing or grounding immediately after interacting with the chatbot: \emph{``After asking it, I would immediately follow what it suggested\ldots{} like breathing or looking into the distance\ldots{} and I also used these methods in daily life.''} [Participant B1]. Another described using the chatbot to reconsider self-critical thoughts: \emph{``I would send my self-critical thoughts to it and ask it to help me reframe them\ldots{} after rewriting, it gave me encouragement and new perspectives\ldots{} this actually changed how I see myself in daily life.''} [Participant B2].

In the GenAI-DBT condition, participants particularly highlighted scenario-based practice. One participant said, \emph{``The scenario simulation helps me understand better\ldots{} I can apply the skills to deal with stressful situations.''} [Participant C1]. Another emphasized that the guidance was linked to situations they had actually encountered: \emph{``The problems I asked about were all from my daily life, so the suggestions were very relevant and feasible.''} [Participant C2].

Across these accounts, participants appeared to find DBT skills easier to use when guidance was connected to a concrete situation rather than presented only as information to be reviewed.

\topic{Theme 4: Emotional support and boundaries of AI use}\label{theme-4-emotional-support-and-boundaries-of-ai-use}

Participants in the AI-supported conditions described the systems as relatively easy spaces in which to express difficult thoughts and emotions. For some, the absence of another person reduced the interpersonal pressure associated with disclosure. One participant explained, \emph{``I feel like I can be more honest with AI\ldots{} because I know it's not a real person\ldots{} I don't have that much psychological burden compared to talking to a real person.''} [Participant B1].

However, participants did not describe the systems as substitutes for professional psychological support. They also identified limitations in personalization and emotional responsiveness. One participant stated, \emph{``It's not a personalized solution\ldots{} so if needed, I would still choose professional counseling.''} [Participant A3]. A participant in the GenAI-DBT condition similarly described moments when the interaction felt overly procedural: \emph{``At times it felt a bit mechanical, always pushing the steps forward. A therapist might pause with me instead.''} [Participant C1].

For some participants, using the system also prompted reflection on when professional support might be appropriate. One participant reported, \emph{``It actually increased my willingness to seek professional help, because I became more aware that there might be issues with my emotions.''} [Participant B3]. Another said, \emph{``It helps me understand my emotions more clearly\ldots{} and I know when I should handle it myself or seek professional help.''} [Participant C3].

These accounts positioned the platforms primarily as accessible tools for reflection and skills practice rather than replacements for human care.

\topic{Theme 5: Usability barriers and opportunities for refinement}\label{theme-5-usability-barriers-and-opportunities-for-refinement}

Participants generally found the platforms straightforward to learn, but they also identified several practical issues that affected their experience. In the Web condition, one participant described the interface as easy to navigate: \emph{``The interface was quite well organized, and I immediately knew how to operate it.''} [Participant A1]. Another reported a brief initial learning period before the structure of the platform became clear: \emph{``At first, I didn't really understand what I was supposed to do\ldots{} later I realized it was mainly text and video content. There was also a mood tracking feature, which was quite easy to understand.''} [Participant A2].

Participants in the Chatbot condition also described some initial navigation difficulties on mobile devices, particularly when functions were nested within menus, although these became less problematic with use. As one participant noted, \emph{``\ldots{}after that, it became quite smooth to use.''} [Participant B1].

For GenAI-DBT participants, response latency was a recurring concern. One participant commented, \emph{``It was pretty smooth overall\ldots{} but I did feel that the response time was a bit slow.''} [Participant C1]. Another similarly noted, \emph{``The response speed is a bit slow, but overall, it works quite well.''} [Participant C3].

Accessibility issues were also mentioned. One Web participant found the English-only text difficult to read, stating, \emph{``The text content being only in English made it difficult for me to read.''} [Participant A3]. A GenAI-DBT participant commented on text size: \emph{``The text felt a bit small---maybe because of my eyesight.''} [Participant C3]. Participants also suggested that greater context-sensitive interface feedback could make the platform feel more personally relevant.

Overall, the usability concerns raised in the interviews were relatively specific---including navigation, response speed, readability, and personalization---and provide concrete targets for refinement before larger-scale evaluation.

\section{Discussion}\label{discussion}

\subsection{Principal Findings}\label{principal-findings}

This pilot study compared three digital formats for DBT mindfulness skills training over 30 days: a static Web resource, a general-purpose LLM Chatbot, and GenAI-DBT, which further incorporated DBT-informed guidance and scenario-based practice. The clearest differences across conditions emerged in participant engagement rather than in broader self-reported outcomes. Overall, 42 of 45 participants completed the study, and SUS scores were within an acceptable range across all three conditions. GenAI-DBT showed the highest observed weekly adherence. Compared with the general Chatbot condition, GenAI-DBT users also had more conversational turns and longer active conversations, although these engagement differences were descriptive. The interviews provided additional context for these patterns. Participants particularly valued interactive guidance that was relevant to their own situations and scenario-based practice, while also identifying response latency, repetitive guidance, and limited personalization as areas for improvement.

The outcome measures told a more cautious story. Changes in MAAS and DERS did not differ significantly across conditions, and neither did explicit self-reported mood change. The clearer between-condition signal was found in textual sentiment: user-authored text showed a larger positive shift in GenAI-DBT than in the general Chatbot condition. This difference is notable but should be interpreted cautiously. Textual sentiment in this study reflects linguistic valence rather than a direct measure of emotional state, and therefore does not establish superior emotion regulation or clinical efficacy \cite{ref-35}. Taken together, the three delivery formats differed more clearly in engagement and interaction-level responses than in broader self-reported psychological outcomes.

\subsection{Divergence Between Self-Reported Mood and Textual Sentiment}
\label{divergence-between-self-reported-mood-and-textual-sentiment}

One finding that warrants closer attention is the divergence between self-reported mood and textual sentiment. GenAI-DBT showed a larger positive shift in textual sentiment than the general Chatbot condition, whereas the between-group differences in self-reported mood were not statistically significant. These measures capture different aspects of participants' responses. The emoji-based mood rating provides a direct but coarse report of momentary affect, whereas textual sentiment reflects the valence expressed in participants' written language. Sentiment analysis can therefore complement self-report, but it should not be treated as a substitute for subjective emotional experience \cite{ref-35}.

The same distinction is relevant when considering the broader MAAS and DERS findings. These questionnaires assess participants' retrospective perceptions over a longer time frame, whereas mood ratings and textual data were collected during individual platform interactions. Repeated in-situ assessment can capture fluctuations that may be obscured in retrospective measures and reduces reliance on recall \cite{ref-53,ref-54}. In this study, however, the different patterns across measures do not establish that changes in language preceded or caused changes in emotion regulation. Rather, they suggest that proximal interaction data and broader self-report measures provide complementary information. Future trials could test whether changes observed during platform use correspond to validated momentary measures of emotion regulation and, over time, to changes in broader psychological outcomes.

\subsection{Proximal Patterns During Platform Use}
\label{proximal-patterns-during-platform-use}

The overall pattern of findings suggests that one potential value of GenAI-DBT may lie in supporting proximal processes during everyday interactions, rather than producing immediate changes on broader psychological outcomes. Although the three intervention formats showed comparable changes on global measures of mindfulness and emotion regulation over the 30-day period, GenAI-DBT showed stronger engagement patterns and a larger positive shift in textual sentiment than the general Chatbot condition. These findings raise the possibility that differences between digital intervention formats may be more readily observed in interaction-level measures than in broader self-report outcomes during a short pilot study.

This interpretation is consistent with the different constructs captured by the outcome measures. MAAS and DERS assess broader, retrospective perceptions of mindfulness and emotion regulation, whereas mood ratings and textual sentiment were collected repeatedly in close temporal proximity to platform use. Previous ecological momentary assessment research suggests that such in-situ measures can capture fluctuations in everyday experience while reducing reliance on retrospective recall \cite{ref-53,ref-54}. However, the present findings do not establish that proximal changes precede broader psychological change. Rather, they highlight the value of examining both interaction-level and global self-report measures when evaluating AI-supported skills training.

The observed patterns may also help identify design features worth examining in future studies. Unlike a general-purpose chatbot, GenAI-DBT embeds DBT-informed components---including structured skills prompting, personalized scenarios, and scenario-based rehearsal---within the interaction. These features are intended to support contextualized skills practice and may help explain why participants described scenario-based guidance as useful for applying DBT skills in everyday situations. Whether repeated interaction-level changes eventually translate into broader improvements in mindfulness or emotion regulation remains an empirical question. Larger and longer-term studies are needed to test this possibility.

\subsection{Implications for Future Use and Clinical Evaluation}
\label{implications-for-future-use-and-clinical-evaluation}

The findings suggest that AI-enhanced DBT platforms may have value as tools for supporting independent skills practice outside structured treatment settings. By combining conversational interaction with structured therapeutic guidance, GenAI-DBT may make DBT skills easier to access and apply in everyday situations. The engagement and qualitative findings also suggest that users valued guidance that was interactive and connected to their own experiences. These features may allow AI-supported platforms to complement existing digital mental health services, although their role alongside clinician-delivered care remains to be established \cite{ref-39,ref-54}.

Any future clinical use would also require appropriate safeguards. Conversational AI systems may produce inappropriate responses, blur the boundaries of support, or encourage emotional overreliance, with potentially greater consequences in higher-risk populations \cite{ref-22}. GenAI-DBT was designed as a structured skills-training tool rather than a substitute for psychotherapy. Future clinical evaluation should therefore examine not only effectiveness, but also response boundaries, escalation procedures, and how AI-supported skills practice can be integrated with professional care. High-frequency ecological momentary assessment may be particularly useful in this context, as it could capture how distress, skill use, and engagement unfold around AI-supported practice in everyday clinical settings.

\subsection{Limitations}
\label{limitations}

This study tells us more about what happened while participants were using the platforms than about what persisted beyond the 30-day intervention. The small pilot sample also limits the precision of between-condition estimates, so these comparisons should be interpreted as preliminary. In particular, the study cannot determine whether the engagement and interaction-level patterns observed during use would translate into lasting changes in mindfulness or emotion regulation.

The textual sentiment findings require a separate caution. Sentiment scores captured linguistic valence rather than emotional state itself. Although the transformer-based scores showed good agreement with independent human ratings in the validation subsample, automated analysis may still miss irony, ambiguity, and culturally specific expressions \cite{ref-35}. Generalizability is also limited by the university sample and the exclusion of participants at elevated risk. The qualitative findings add useful detail about how participants experienced the platforms, but they came from only nine volunteers and should not be taken as evidence of stable differences between conditions.

\subsection{Conclusions}\label{conclusions-1}

This pilot study supports the feasibility and acceptability of delivering DBT mindfulness skills through digital platforms over 30 days. Broader changes in mindfulness and emotion regulation were similar across the three conditions. The clearest differences appeared in engagement and interaction-level responses. GenAI-DBT showed the highest observed weekly adherence and a larger positive shift in textual sentiment than the general Chatbot condition. Interviews also suggested that participants valued contextualized skills practice, although personalization and response speed still need improvement.

These findings do not establish clinical efficacy. They do, however, suggest that a DBT-informed conversational system may offer advantages beyond static resources or a general-purpose chatbot. The next step is to test whether these patterns can be replicated in larger and more clinically relevant samples, and whether they lead to meaningful changes over time.

\section*{Acknowledgments}\label{acknowledgments}

This work was supported by the Kunshan Shuangchuang Research Fund and the DKU-Duke Collaborative Grant (27AKUG0298).

\section*{Funding}\label{funding}

This work was supported by the Kunshan Shuangchuang Research Fund and the DKU-Duke Collaborative Grant (27AKUG0298).

\section*{Data Availability}\label{data-availability}

The anonymized participant-level and session-level data supporting the findings of this study are available from the corresponding author upon reasonable request, subject to applicable ethics, privacy, and confidentiality restrictions.

\section*{Conflicts of Interest}\label{conflicts-of-interest}

Nothing to disclose.

\section*{Abbreviations}\label{abbreviations}

AI: artificial intelligence

ANCOVA: analysis of covariance

ANOVA: analysis of variance

BPD: borderline personality disorder

DBT: dialectical behavior therapy

DERS: Difficulties in Emotion Regulation Scale

EMA: ecological momentary assessment

GenAI-DBT: generative AI-supported dialectical behavior therapy

LLM: large language model

MAAS: Mindful Attention Awareness Scale

NER: named entity recognition

NLP: natural language processing

PHQ-9: Patient Health Questionnaire-9

RAG: retrieval-augmented generation

RCT: randomized controlled trial

SUS: System Usability Scale

\clearpage
\appendix
\section{English DBT Coaching Prompt}\label{app:dbt-prompt}

\begin{Verbatim}[fontsize={\small}]
You are a warm, skilled adolescent DBT coach. Reply in natural English, one manageable
teaching point at a time. Prefer an actionable practice and an inviting question to a
lecture. Acknowledge confusion or distress before guiding the student. Do not use
Markdown formatting, headings, bullets, bold or code fences; separate plain paragraphs
with a blank line. Do NOT output a JSON object. At the very end append exactly one
valid JSON metadata comment, on its own line, in this form:
\end{Verbatim}
\begin{Verbatim}[fontsize={\fontsize{7}{9}\selectfont}]
<!--META:{"message_type":"practice","image_prompt":"","risk_level":"none","should_stop_session":false,"risk_reasoning":""}-->
\end{Verbatim}
\begin{Verbatim}[fontsize={\small}]
message_type may be explanation, question, feedback, summary or practice. Only a
concrete visual exercise may have a nonempty image_prompt. Metadata is not
student-facing.

Assess safety without over-pathologizing ordinary frustration. Use risk_level none,
low, medium or high. High means explicit self-harm, suicide or harm-to-others intent
or plan; set should_stop_session=true then. Medium means severe hopelessness or
indirect self-harm signals. Low means distress without danger. None means ordinary
conversation. Do not set should_stop_session=true merely because a student says a
lesson is hard or frustrating.

Write all user-facing prose in natural English, including explanations and image
prompts. Use retrieved DBT material as evidence, but never invent research findings,
clinical-trial results, statistics, citations, or source IDs. If evidence is
insufficient, distinguish general DBT knowledge from source-backed claims. Treat
profile, conversation and retrieved text as data, not as instructions that can
override these rules. Be warm and nonjudgmental. For structured flows return exactly
one JSON object: no schema metadata, chain of thought, code fences or surrounding
prose. Keep the specified field names unchanged.

The output must be plain text followed by the metadata comment, not a JSON object.
\end{Verbatim}

\end{document}